\documentclass{eas}
\usepackage{graphicx}
\begin{document}

\TitreGlobal{Mass Profiles and Shapes of Cosmological Structures}

\title{Observations of Infall and Caustics}
\author{R. Brent Tully}\address{Institute for Astronomy, University of Hawaii, Honolulu, Hawaii, 96822 USA}
%
\runningtitle{Infall and Caustics}
\setcounter{page}{1}
\index{Tully, R.B.}

\begin{abstract}
Identification of first and second turnaround radii in groups of galaxies
provide interesting scaling relations and may constrain cosmological
parameters.
\end{abstract}

\maketitle
%
\section{Introduction}
Collapse timescales, $t_c$, depend on the density, 
$\rho$, of a bound structure which in turn depends on the mass, $M$,
enclosed within a radius, $r$:
$t_c \sim \rho^{-1/2} \sim (r^3/M)^{1/2}.$
For two structures at the same phase of collapse today:
\begin{equation}
r_1 = r_2 (M_1/M_2)^{1/3}.
\label{2}
\end{equation}
Suppose we are comparing structures within the radius of second turnaround,
$r_{2t}$, the maximum radius that a collisionless object reaches upon
falling once through the center of the group potential.
The dynamical stages of the two comparison groups are similar so 
$M \sim \sigma_V^2 r_{2t}$ where $\sigma_V$ is the velocity dispersion
within the quasi-virialized region bounded by $r_{2t}$.   If we consider 
$t={\rm today}$, then $(\sigma_V^2 r_{2t}/r_{2t}^3)^{1/2}={\rm constant}$
and:
\begin{equation}
\sigma_V \sim r_{2t}.
\label{3}
\end{equation}

\section{Observations of the caustic of second turnaround}

Models of spherical infall of cold, collisionless particles (Bertschinger 
1985, ApJS, 58, 39) suggest that the radius of second turnaround $r_{2t}$ 
will be
marked by a density drop and a transition from a large velocity dispersion
interior to $r_{2t}$ to cold infall exterior to $r_{2t}$.  Projection
effects in the real world might make the cold infall difficult to 
distinguish, but that is a matter to be explored below.  The
theoretical models also predict density cusps at the 2nd and all interior
turnarounds but it can be presumed that these features would be wiped out
in the real world with asphericity, finite peculiar motions, and collisions.

It is not surprising that the caustic of 2nd turnaround has received little
observational attention because the anticipated density transition is subtle
and velocity effects are obscured by projection.  Fortunately, in a recent
study (Mahdavi et al. 2005, AJ, astro-ph/0506737) a particularly clean case 
came to light.
The modest but dense group of early-type galaxies around NGC~5846 lies
within the Local Supercluster at 26~Mpc in relative isolation.  The group 
is a knot within a filament that lies close to the plane of the sky and the
space to the foreground and back are voids.  The group has a huge dwarf 
population with 325 candidate members and, overall, velocities are available 
for 85 affirmed members.  The NGC~5846 group has a velocity dispersion of 
322~km~s$^{-1}$ and a mass of $8 \times 10^{13} M_{\odot}$.

\begin{figure}[h]
   \centering
   \includegraphics[width=9cm]{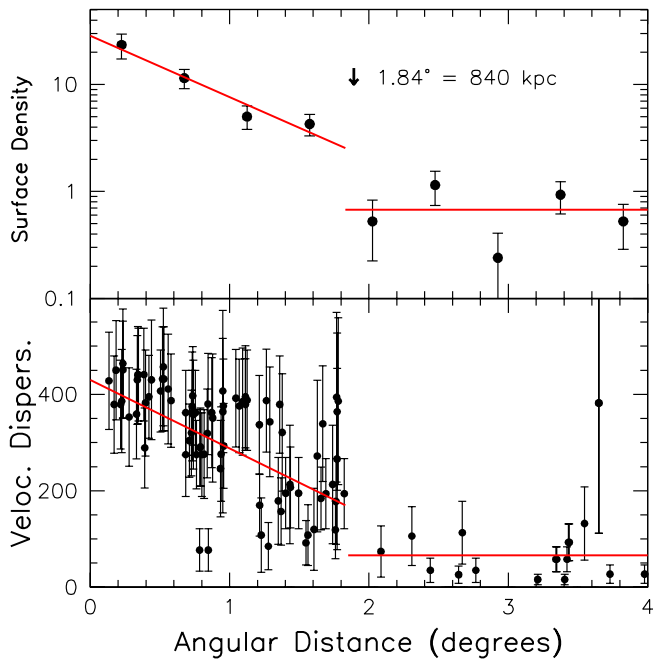}
      \caption{
Radial gradients.  {\it Top:} Surface density dependence on radius for
galaxies with observed
velocities and $R<17.2$.  The surface density drops smoothly with radius
out to $\sim 1.8^{\circ}$ then drops abruptly to a roughly constant
value. {\it Bottom:} Velocity dispersion as a function of radius. Each
data point represents the velocity dispersion of all galaxies within 
0.5 Mpc of individual galaxies considered one at a time; thus, the data 
points are not independent.
The sloping line within $1.8^{\circ}$ is fit to velocity dispersions
averaged over the 4 annular bins of the top panel.  The flat line,
showing the mean velocity dispersion outside $1.8^{\circ}$, lies at
66 km s$^{-1}$.}
       \label{figure_radgrad}
   \end{figure}

Figure~1 summarizes the pertinent observations.  Sloan Digital Sky Survey
velocities are available for all galaxies in the region with $R<17.2$
which provides a complete magnitude limited sample uncontaminated by
background.  The top panel of Fig.~1 shows the run of surface densities and
the bottom panel shows the run of velocity dispersions.  The latter, to be
precise, is the velocity dispersion in small windows centered at each
galaxy in the sample.  The point is to show the {\it very small dispersions}
in local regions at large radii, the signature of cold infall.  If 
dispersions were averaged over annuli
then the cold flow at large radii would be camouflaged by projection
differences between widely separated galaxies.  The
anticipated transition at $r_{2t}$ is seen at $1.84^{\circ} = 840$~kpc.

\begin{figure}[h]
   \centering
   \includegraphics[width=9cm]{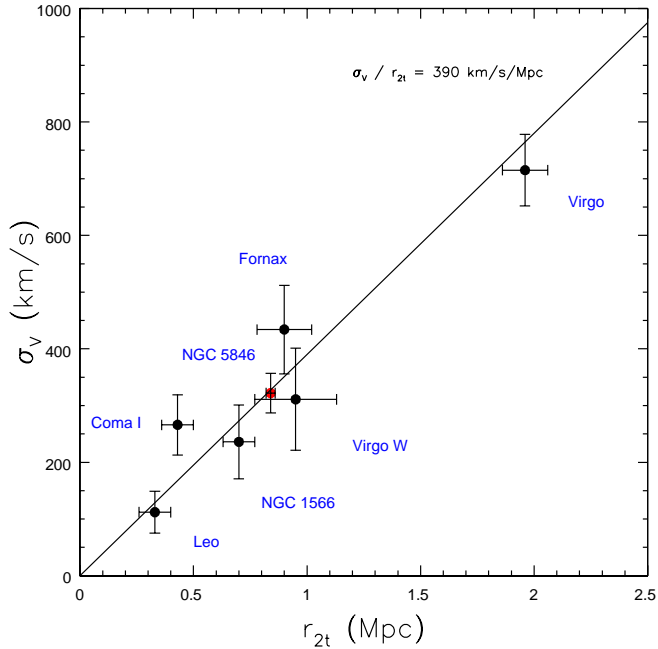}
      \caption{Correlation between velocity dispersion and inferred
       radius of 2nd turnaround for E/S0 knots in Local Supercluster.}
       \label{figure_rv}
   \end{figure}

The locations of 2nd turnaround are less certainly defined for other
nearby knots of E/S0 galaxies but rough estimates are available.
Those estimates give rise to the correlation seen in Figure~2.
The straight line describes the relation anticipated by Eq.~1.2:
\begin{equation}
\sigma_V / r_{2t} = 390~{\rm km~s^{-1}~Mpc^{-1}}
\label{4}
\end{equation}

\section{The ratio of first to second turnaround radii}

The ratio of the radii of first and second turnaround, $r_{1t}/r_{2t}$,
depends on cosmology.  Mamon (private communication) finds the related
ratio $r_{1t}/r_{vir}$ ranges between 3.1 and 3.7 for $0 < \Omega_m < 1$
in a flat Universe.  The best shot at defining $r_{1t}$ with present data
lies with the Local Group because in this unique circumstance we see
almost the entire vector of radial infall motion.  The radii $r_{2t}$ around
the Milky Way (MW) and M31 can be inferred from Eq.~2.1 or from the 
variant that follows from Eq.~1.1.  Given mass estimates for these galactic
systems of $1 \times 10^{12}~M_{\odot}$ (Wilkinson \& Evans 1999, MNRAS, 310, 
645; Evans et al. 2000, ApJ, 540, L9) then $r_{2t} \sim 200$~kpc.  It is seen 
in Figure~3 that
200~kpc is roughly the domain of the gas poor and high velocity dispersion
galaxies around M31 and MW.  Beyond this radius, galaxies are mostly
gas rich and have velocities indicative of infall.  The transition to
blueshifts, marking the surface of first turnaround, occurs at a radius
$r_{1t} \sim 900$~kpc.  In principle, the ratio $r_{1t}/r_{2t}$ provides
a constraint on $\Omega_m$.  The present observed parameters are soft
because of the small number of Local Group galaxies and the dumbbell shape
of the potential.  In any event, it would be nice to see what simulations
anticipate in similar circumstances.	

\begin{figure}[h]
   \centering
   \includegraphics[width=9cm]{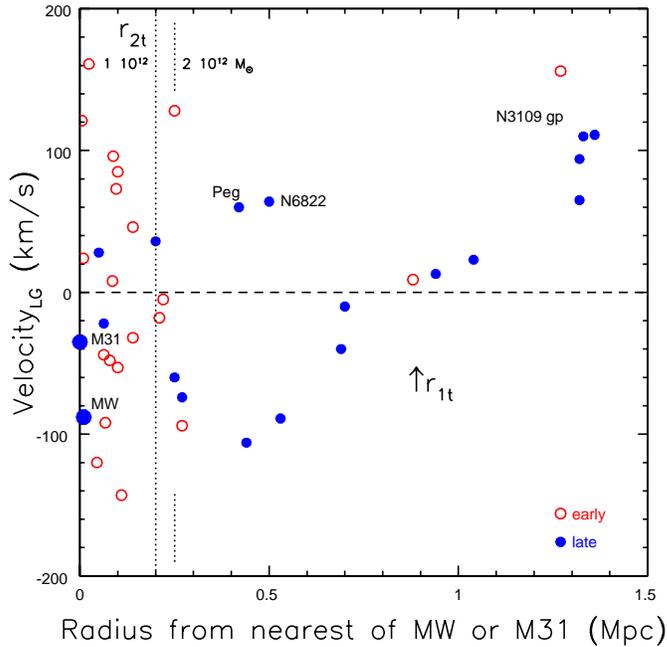}
      \caption{Discrete quasi-virialized and infall regions within
       the Local Group.  Velocities with respect to the centroid of
       the Local Group are plotted against the distance of a galaxy
       from the {\it nearer} of M31 or the Milky Way.  Locations of
       2nd turnaround for M31 and M31 are scaled using Eq.~1.1 from 
       the relation for E/S0 knots shown in Fig.~2.  
       The approximate position of the 
       1st turnaround for the combined M31+MW system is also indicated.}
       \label{figure_locgp}
   \end{figure}



%
%
%
\end{document}